\documentclass[journal,twoside,web]{ieeecolor}
\usepackage[dvipsnames]{xcolor}
\usepackage{generic}
\usepackage{amsmath,amssymb,amsfonts}
\usepackage{algorithmic}
\usepackage{graphicx}
\usepackage{textcomp}
\usepackage{url}
\usepackage{academicons}
\usepackage{array}
\usepackage[style=numeric,sorting=none]{biblatex}
\usepackage{subfigure}
\begin{document}
\title{Ultrahigh Frequency and Multi-channel Output in Skyrmion Based Nano-oscillator}
\author{
  Abhishek Sharma, %\orcid{0000-0002-3635-3081} \and
  Saumya Gupta, %\orcid{0009-0001-7636-1553} \and
  Debasis Das, %\orcid{0000-0002-4293-5838} \and
  Ashwin. A. Tulapurkar, %\orcid{0000-0002-6539-5571} \and
  Bhaskaran Muralidharan %\orcid{0000-0003-3541-5102}
\thanks{Abhishek Sharma and Saumya Gupta are co-first authors. }
\thanks{ Abhishek Sharma is with the Department of Electrical Engineering, IIT
Ropar, Rupnagar, Punjab 140001, India (e-mail: abhishek@iitrpr.ac.in). }
\thanks{Saumya Gupta, Bhaskaran Muralidharan, and  Ashwin. A. Tulapurkar are with the Department of Electrical Engineering, IIT Bombay, Mumbai 400076, India
(e-mail: saumyaguptaalld@gmail.com; bm@ee.iitb.ac.in;ashwin.tulapurkar@iitb.ac.in).}
\thanks{The author BM acknowledges the support by the Science and Engineering Research Board (SERB), Government of India, Grant No. CRG/2021/003102 and Grant No. MTR/2021/000388, and the Ministry of Human Resource Development (MHRD), Government of India, Grant No. STARS/APR2019/NS/226/FS under the STARS scheme.  }
\thanks{Debasis Das is with the Department of Electrical and Computer
Engineering, National University of Singapore, Singapore 117583
(e-mail: eledd@nus.edu.sg).}}
\maketitle

\begin{abstract}
Spintronic nano-oscillators can generate tunable microwave signals that find a wide range of applications in the field of telecommunication to modern neuromorphic computing systems. Among other spintronic devices, a magnetic skyrmion is a promising candidate for the next generation of low-power devices due to its small size and topological stability. In this work, we propose a multi-channel oscillator design based on the synthetic anti-ferromagnetic (SAF) skyrmion pair. The mitigation of the skyrmion Hall effect in SAF and the associated decimation of the Magnus force endows the proposed oscillator with an ultra-high frequency of 41GHz and a multi-channel frequency output driven by the same current. The ultrahigh operational frequency represents an $\sim$342 times improvement compared to the monolayer single skyrmion oscillator featuring a constant uniaxial anisotropy profile. Using micromagnetic simulations, we demonstrate the effectiveness of our proposed multi-channel oscillator design by introducing multi-channel nanotracks along with multiple skyrmions for enhanced frequency operation. The ultrahigh operational frequency and multi-channel output are attributed to three key factors: The oscillator design accounting for a finite spin-flip length of the spacer (such as Ru) material, tangential velocity proportionality on input spin current along with weak dependence on the radius of rotation of the skyrmion-pair, skyrmion interlocking in the channel enabled by the multi-channel high Ku rings and skyrmion-skyrmion repulsion, therefore resulting ultrahigh frequency and multi-channel outputs.
\end{abstract}

\begin{IEEEkeywords}
Bilayer, gigahertz frequency, nano oscillators, skyrmions, spintronics, Synthetic antiferromagnet, Spin Transfer Torque
\end{IEEEkeywords}

\section{Introduction}
\label{sec:introduction}
\IEEEPARstart{T}{he} spin-transfer torque (STT)\cite{slonczewski1996current}, along with magnetoresistance (MR)\cite{butler2001spin}, has emerged as a technologically relevant phenomenon that caters to various applications such as spin transfer torque-magnetic random access memory (STT-MRAM)\cite{gajek2012spin}, spin-transfer torque nano-oscillators
(STNOs)\cite{deac2008bias}, racetrack memories\cite{parkin2008magnetic}, to name a few. The STNOs are a class of non-linear inductorless oscillators in which the STT excites uniform magnetization dynamics, and non-collinear MR translates the precession of magnetization to the microwave signal\cite{sharma2017resonant}. Among various spintronic devices, magnetic skyrmion show a huge potential for next-generation memory and computing devices, due to its small size \cite{wang2018theory}, topological protection\cite{Skyrmion_soccer}, low depinning current density\cite{Romming636}, electrical manipulation and detection\cite{Sampaio2013}. Thus, making them suitable candidate for many applications such as nano-oscillator\cite{garcia2016skyrmion}, race track memories\cite{Tomasello2014}, logic gates \cite{zhang2015magnetic,luo2018reconfigurable}, neuromorphic computing\cite{huang2017magnetic,bhattacharya2019low,das2023bilayer} and transistors\cite{Zhang2015a}. The observation of magnetic skyrmions in a bulk non-centrosymmetric ferromagnet (FMs)\cite{Skyrmionlattice2009,Jonietz1648,Yu2010} and thin films\cite{Heinze2011,Hrabec2017} has paved a way for many advanced electronic applications. The magnetic skyrmions are stable vortex-like spin textures with a non-trivial topology. The magnetic skyrmions are stabilized by a mutual competition between Heisenberg exchange interaction favouring collinear alignment and Dzyaloshinskii-Moriya interaction (DMI) interaction favouring perpendicular alignment of spins \cite{fert2013skyrmions}. Bulk non-centrosymmetric magnets with bulk DMI support the Bloch-type magnetic skyrmions, whereas in thin films, Neel-type hedgehog skyrmions are stabilized by the interfacial DMI acting on the FM layer attached to a heavy metal (HM) layer \cite{Nagaosa2013,Fert2017}. It has been shown that a thin FM layer with a perpendicular magnetic anisotropy (PMA) grown on a heavy-metal layer, can stabilize Neel-type Skyrmion\cite{Heinze2011,Hrabec2017}. Robustness against the thermal fluctuation \cite{kang2016skyrmion, wang2022single}, makes the Neel-type of skyrmion, one of the promising candidates for future spintronic applications. Thus, in this work, we focus on the Neel-type skyrmion to design the spintronic oscillator. \\ 
\indent In particular, the skyrmion-based oscillators\cite{garcia2016skyrmion,zhang2015magnetic,guo2021ferromagnetic,guo2020ferromagnetic,jin2020high} have been explored theoretically for microwave generation featuring a small line-width and dynamical frequency range (MHz $\sim$ few GHz) reporting manipulation of PMA, DMI, polarization angle, or design level variations such as introducing a groove in the ferromagnet layer to enhance operational frequencies. Jin et. al. reported an operation frequency of 15.63GHz by moving a skyrmion in an annular groove\cite{jin2020high}. Our proposed skyrmion STNO is feasible for introducing multiple skyrmions together to improve the oscillation frequency, having both a smaller size and topological protected dynamics compared to regular STNOs\cite{garcia2016skyrmion,zhang2015current,Das2019}. The main obstacle faced by the skyrmion is the Magnus force\cite{stone1996magnus} which adds a transverse component to the velocity, leading to a path deviation governed by the injected spin current, a phenomenon known as the skyrmion Hall effect\cite{Skyrmion_soccer}. This may lead to skyrmion annihilation at the edge of the FM layer for large spin currents, thus limiting the gyration frequency of a skyrmion-based oscillator\cite{garcia2016skyrmion,Das2019}. To resolve this issue of the unwanted Magnus force, we utilize the bilayer system, where two FM layers are coupled by an antiferromagnetic (AFM)\cite{zhang2016thermally,legrand2020room,juge2022skyrmions} exchange coupling\cite{barker2016static}. In such a system, sufficient coupling constant ensures the cancellation of the Magnus force acting on the skyrmions in the individual FM layers\cite{zhang2016magnetic}.\\
Several other variations of skyrmion-based oscillators utilizing the AFM or bilayer geometry\cite{shen2019spin} as reported by Shen et. al. depicted an AFM system with skyrmion having a frequency of 25GHz by varying the nanodisk radius, damping constant, and applied current\cite{shen2019spin}.\\
\indent In this work, we explore a skyrmion-based oscillator design using an AFM exchange-coupled bilayer system. Our work underscores the significance of our design in addressing a limitation often overlooked in prior studies involving bilayer skyrmion geometries, namely, the substantial spin-flip length associated with Ru\cite{Ru_spinflip_lenght}. Consequently, our work yields a notably improved upper-limit operational frequency compared to earlier reported works.

Using micromagnetic simulation, we show that our proposed bilayer system can generate a broad range of frequencies compared to its single-layer FM counterpart. We also show that the frequency range can be enhanced by designing a multi-channel circular racetrack, which acts as a multi-frequency generating oscillator using a single device. Using the repulsion characteristics between skyrmions with the same topology, we simulate the device by nucleating multiple skyrmions in the racetracks and demonstrate that the generated frequency range can be increased further. The rest of the paper is organized as follows. In Sec.\ref{Device_model}, we describe our proposed bilayer model for the skyrmion oscillator. Sec.\ref{math_model} describes the mathematical framework used for the simulation. In Sec. \ref{results}, we discuss the results obtained from the simulation, and finally, Sec. \ref{conclusion} concludes the paper.

\section{Device schematic and working principle}\label{Device_model}

\begin{figure}[!t]
\centering
\centerline{\includegraphics[width=1\columnwidth]{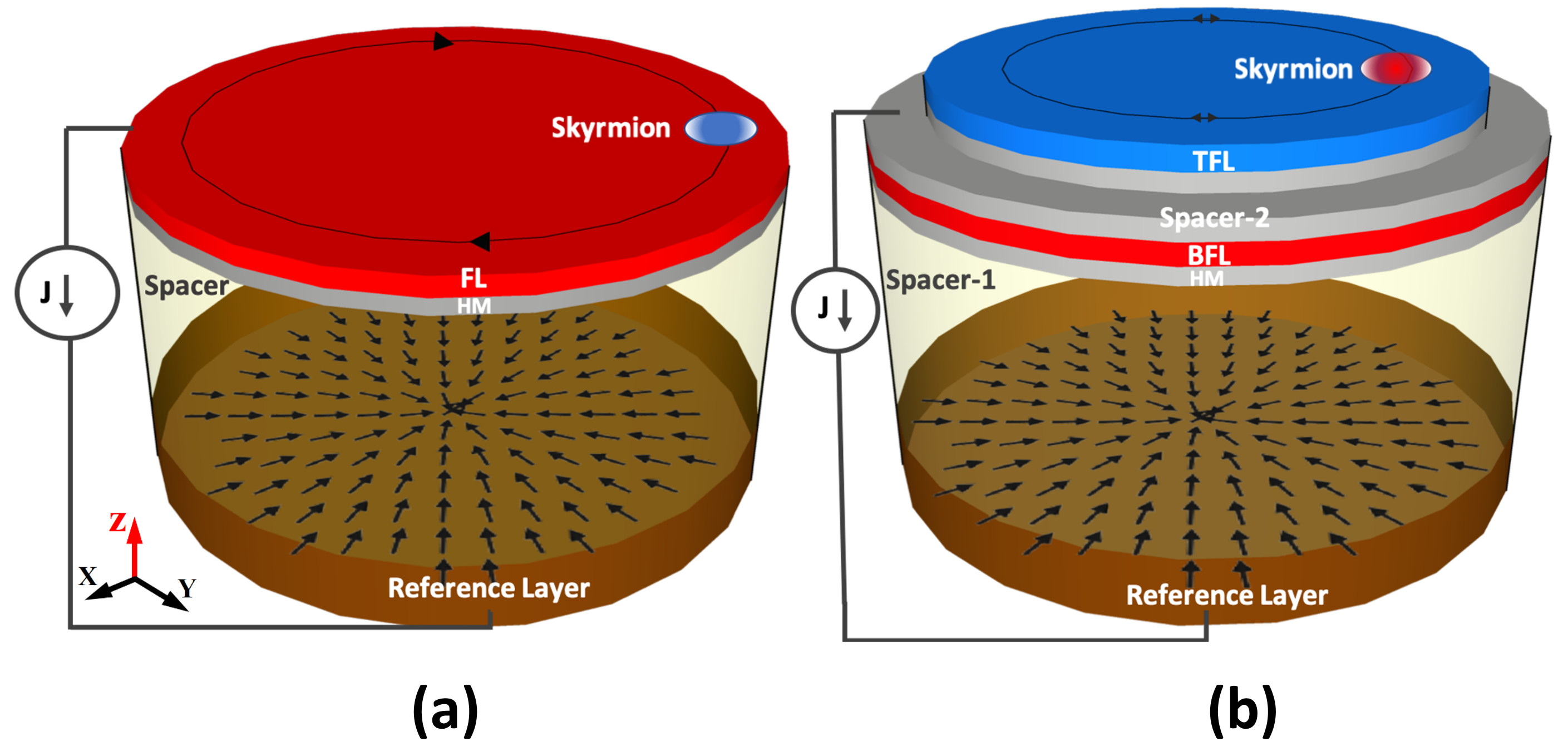}}
\caption{Schematic diagram of (a) regular skyrmion oscillator and (b) bilayer skyrmion based oscillator.}
\label{oscillator_schematic}
\end{figure}
\begin{figure}[!t]
	\centering
\centerline{\includegraphics[height=1.6in,width=2.1in]{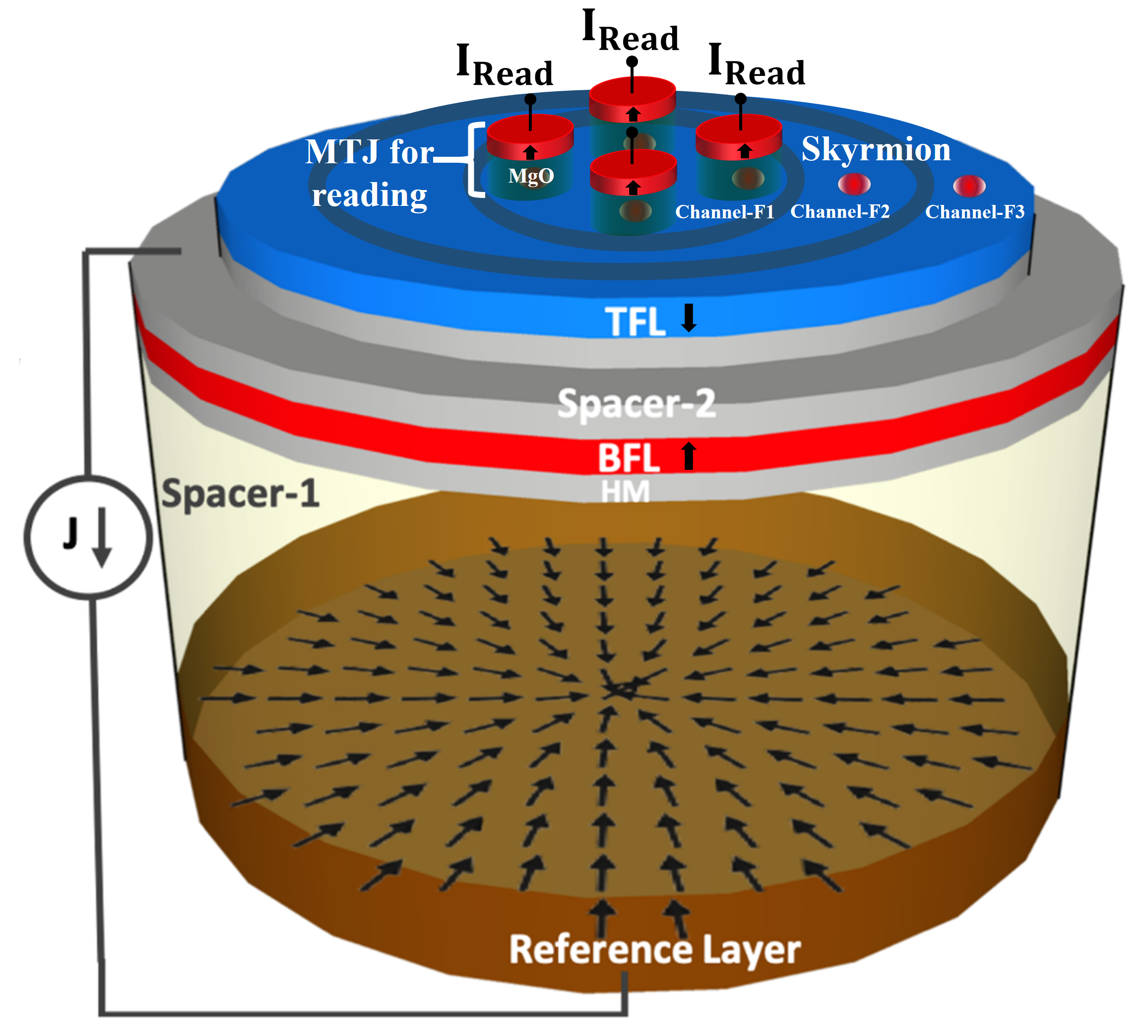}}
	\caption{Schematic diagram of Multi-channel skyrmion-based bilayer nano oscillator with multiple MTJ to detect the corresponding skyrmions present in channel F1(see Fig.\ref{bilayer_skyrmion_oscillator_multi_channel})}
	\label{skyrmionundermtj.png}
\end{figure}

 
\indent A regular skyrmion-based oscillator\cite{garcia2016skyrmion}, consists of two circular FM layers known as the reference layer (RL) and free layer (FL) separated by an insulator layer (spacer) as shown in Fig. \ref{oscillator_schematic}(a). The FL is assumed to have an uniaxial perpendicular magnetic anisotropy (PMA), which is grown on a heavy metal (HM) layer. This induces DMI at the interface of the HM/FM structure due to the broken inversion symmetry. A skyrmion is nucleated in the FL, where the competition between the DMI and Heisenberg exchange interaction in the FL stabilizes magnetic skyrmion \cite{fert2013skyrmions}. On the other hand, the magnetization of the RL is fixed for this structure with in-plane vortex-like spin polarization as shown in Fig. \ref{oscillator_schematic}(a). The vortex-like RL magnetization is denoted by  $\mathbf{m}_{\text{RL}}=(cos\phi,sin\phi,0)$, where $\phi=tan^{-1}(\frac{y}{x})+ \Psi$ and $(x,y)$ are the spatial coordinates of the RL with the origin being at the center of nano-disk(dot) \cite{phatak2012direct,wintz2013topology} and the polarization angle $\Psi$ is $180^{o}$. During the injected current flow, electrons passing through the RL get spin-polarized according to $\mathbf{m}_{\text{RL}}$. The spin-polarized electrons upon reaching the FL, move the skyrmion in a circular path as shown in Fig. \ref{oscillator_schematic}(a). The vortex-like spin-polarized current exerts a tangential force on the skyrmion (see Eq.~\ref{F_t_simplifed}), which causes the motion of the skyrmion in the tangential direction. On the other hand, the Magnus force proportional to the skyrmion velocity (see Eq.~\ref{thiele_2a}) acts along the radially outward direction. This force is balanced by the repulsion force on the skyrmion acting towards the center of the nano-disk due to the canted spins at the edge of the nano-disk\cite{zhang2015skyrmion}. This causes the skyrmion to move in a uniform circular path. The uniform circular motion of the skyrmion can be translated to the electrical signal by the MR effect of a sensing magnetic tunnel junction (MTJ) on top of the FL\cite{kang2016skyrmion}. we also propose an extended schematic design Fig.\ref{skyrmionundermtj.png} for reading via Magnetoresistance in light of recent experimental demonstration\cite{li2022experimental,penthorn2019experimental}, the proposed device utilizes multiple synchronized MTJ's corresponding to each skyrmion in a particular channel Fig.\ref{bilayer_skyrmion_oscillator_multi_channel} which can be phase-locked to deliver an enhanced additive output power from each MTJ compared to a single MTJ to read the nanodisk. The frequency of the skyrmion oscillator increases with current density (J) attributed to the rise in skyrmion tangential velocity. The intensified Magnus force as an effect of increased skyrmion velocity can surpass edge repulsion beyond a critical current density value leading to skyrmion annihilation at the edge of the nano-disk\cite{Das2019}. This limits the gyration frequency of the skyrmion oscillator as well as any possibility of multichannel operation.\\
\indent To alleviate this issue, we explore a bilayer device for the skyrmion-based oscillator as shown in Fig.~\ref{oscillator_schematic}(b). It has a similar device design to the regular skyrmion-based oscillator, except for the free layer. In the bilayer device, a single free ferromagnet layer is replaced by two free ferromagnet layers known as the bottom free layer (BFL) and the top free layer (TFL) which are separated by another spacer layer (spacer-2). The BFL and TFL have the same uniaxial PMA with magnetization pointing along $+\hat{z}$(red) and $-\hat{z}$(blue) directions, respectively. These layers are anti-ferromagnetically coupled (AFC) via spacer-2 consisting of a non-magnetic metal such as Ru through Ruderman-Kittel-Kasuya-Yosida(RKKY) type interlayer coupling\cite{Yang2015}. 
 Additionally, the Ru also provides DMI for stabilizing the skyrmion in the SAF multilayer. The experimental feasibility of such structures in nucleating stable skyrmions is demonstrated by Juge et. al\cite{juge2022skyrmions}.\\We have made the spacer-2 layer such that it touches both the BFL and TFL for different disc radii i.e., a lesser disc surface for TFL compared to the BFL to limit or minimize the spin current flow in the TFL. The design modification(etched design) ensures the application of higher current density in the BFL, and the spacer-2 ensures reduced spin polarization within the SAF multilayer, also accounting for the Ru's spin-flip length of 14nm attributing to higher frequency generation.  if the spin
current translates into the TFL the magnetization dynamics would be completely different, and the skyrmion in the TFL and BFL may no longer be in unison, undermining the objective of the bilayer structure.\\
We have made contact inside the spacer-2 layer touch the BFL (Fig~\ref{oscillator_schematic}(b)) to avoid or minimize the spin current flow in the TFL \cite{Ru_spinflip_lenght}. Without the integration of this design modification, the bilinear surface exchange coefficient reduces beyond the spin-flip length of 14nm for the Ru material leading to a spin memory loss. Under high current densities, the skyrmion in the TFL and BFL will no longer be in unison, undermining the objective of a bilayer structure. 

This type of contact can be made via selectively etching out the section of TFL along with a significant part of the spacer-2 layers. This technique is usually employed in heterostructure-based device fabrication and is well within current fabrication capabilities\cite{BanerjeeD}.\\
\begin{figure}[!t]
	\centering
 \centerline{\includegraphics[width=1\columnwidth]{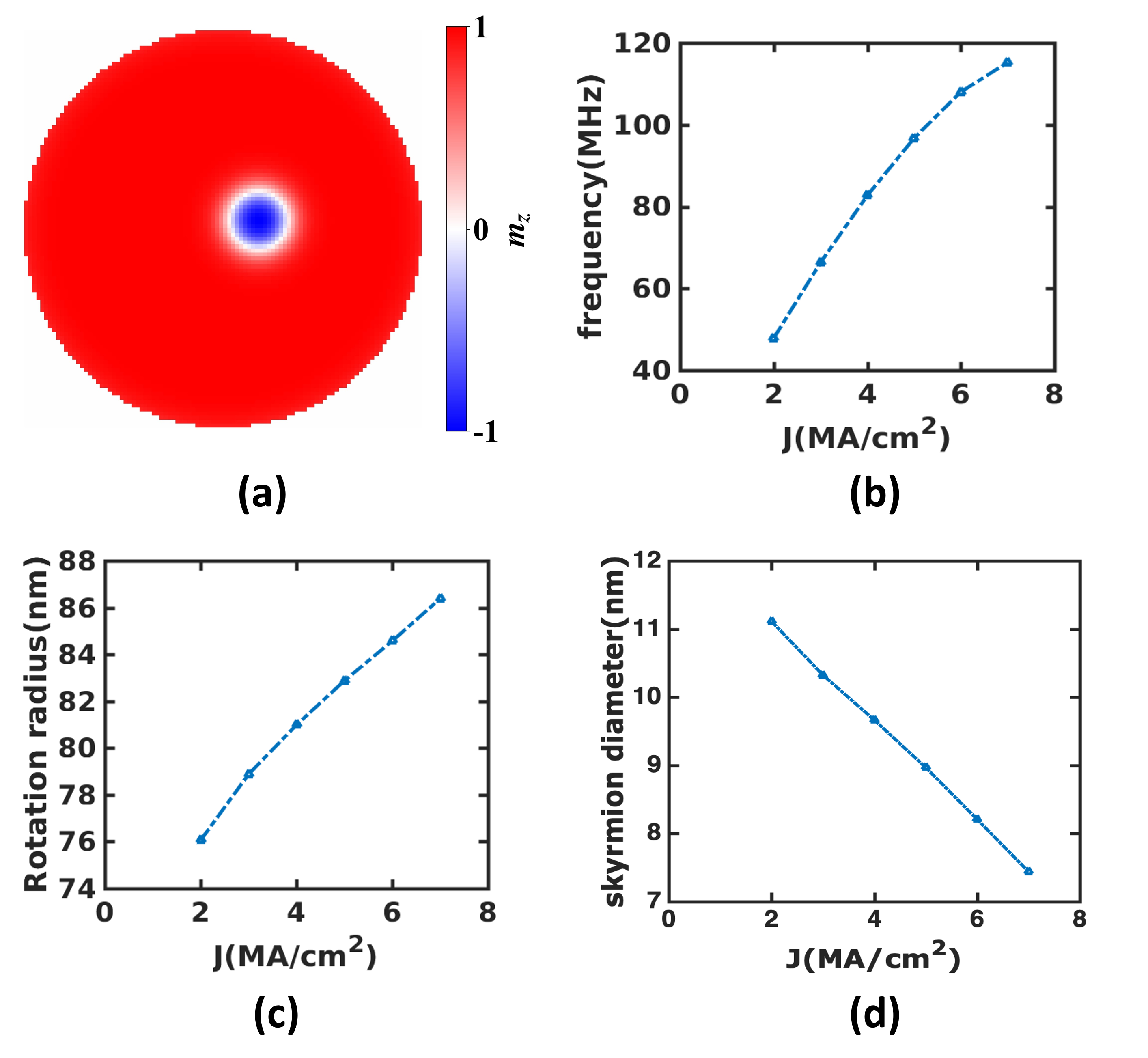}}
	\caption{Monolayer skyrmion oscillator. (a) The two-dimensional color plot of the $m_z$ ($z$-component magnetization) on the $x-y$ plane of the FL. (b) The frequency of oscillation (c) The rotation radius of skyrmion on nano-dot (d)skyrmion diameter with current density. }
	\label{skyrmion_oscillator_chracterstics}
\end{figure}
Along with the monolayer and bilayer skyrmion oscillators with constant PMA we have simulated the bilayer oscillators with other uniaxial anisotropy profiles such as a high ku selective disc and multichannel ring, as described in sec IV can be fabricated using He$^+$ ion irradiation technique\cite{juge2021helium} or VCMA methods which can create desired PMA profile via inducing voltage on the electrode gate through an insulator\cite{xia2018skyrmion}.

\indent The bilayer skyrmion-based oscillator works similarly to the regular skyrmion oscillator except for the absence of Magnus force. The TFL and BFL skyrmions form an AFC skyrmion pair with equal and opposite skyrmion number\cite{zhang2016magnetic,zhang2016thermally}. The Magnus force on the skyrmion pair is zero due to the vanishing gyro-vector(Eq.~\ref{thielebi_eq}). The vertically injected vortex-like spin-polarized current in the BFL exerts a tangential force on the BFL-skyrmion and due to the AFC, both the skyrmions in BFL and TFL move in unison along a uniform circular path as shown in Fig.~\ref{oscillator_schematic}(b). The spin current only exerts a tangential force(Eq.~\ref{thielebi_b}) on the skyrmion pair, which enables a large spin current ejection and high frequency of oscillation. The absence of Magnus force allows the skyrmion pair to move in a designated track opening up the possibility of multi-channel frequency of operation. The uniform circular motion of the skyrmion-pair can be translated to the electrical signal by the MR effect of a sensing magnetic tunnel junction (MTJ)\cite{hanneken2015electrical,zhang2018skyrmions,penthorn2019experimental,li2022experimental} on top of the TFL\cite{kang2016skyrmion} in the path of TFL-skyrmion.

\section{Mathematical description and simulation details}
\label{math_model}
The Hamiltonian of the layer $l$ is given by:
\begin{eqnarray}		
	H_l&=&-J_{ex}\displaystyle\sum_{<i,j>} \textbf{m}_i^{l} \cdot \textbf{m}_j^{l} + \displaystyle\sum_{<i,j>} \textbf{D}_{i,j} \cdot (\textbf{m}_i^{l} \times \textbf{m}_j^{l})\nonumber \\
	&& + K_u \displaystyle\sum_{i}[1-(\textbf{m}_i^l\cdot\textbf{z})^2] + H_{DD} 			
\end{eqnarray}
where, $l$ represent the layer index ($l$=TFL, BFL), $\textbf{m}_i=\textbf{M}/M_s$ is the normalized local magnetic moment at lattice site $i$, $<i,j>$ runs over neighboring sites and $M_s$ is the saturation magnetization. The first term represents the FM exchange interaction ($J_{ex}>0$). The second term represents the DMI interaction with $\textbf{D}$ being DMI vector favoring canted magnetic moments. The third represents uniaxial PMA with $K_u$ being the anisotropy constant and the last term represents the dipole-dipole interaction. Apart from $H_{\text{TFL}}$ and $H_{\text{BFL}}$, there is also another energy term that represents the interlayer exchange coupling between TFL and BFL, given by:
\begin{equation}
	H_{\text{interlayer}}=A_{\text{inter}}\displaystyle\sum_{i} \textbf{m}_i^{\text{TFL}} \cdot \textbf{m}_i^{\text{BFL}}
\end{equation}
where, $A_{\text{inter}}<0$ represents the interlayer AFM interaction. The total Hamiltonian for the bilayer system is $H_T=H_{\text{TFL}}+H_{\text{BFL}}+H_{\text{interlayer}}$. In the continuous limit of the magnetization, the total energy can be written as:

\begin{eqnarray}		
	E&=&\int dV \big[ \big. A(\nabla \textbf{m})^2 + \varepsilon_{DMI} + K_u\left[1-(\textbf{m}\cdot\textbf{z})^2\right]\nonumber \\
	&& - \frac{\mu_0}{2} \textbf{m} \cdot \textbf{H}_d + \varepsilon_{interlayer}\big. \big]		
\end{eqnarray}	
where, A is exchange stiffness constant, $H_d$ is demagnetization field, $\mu_0$ is the free space permeability, $\epsilon_{DMI}$ and  $\epsilon_{interlayer}$ are interfacial form of DMI and interlayer AFM coupling energy densities, respectively and are given by:
\begin{equation}
	\varepsilon_{interlayer}=\sigma \left[1-\textbf{m}^{TFL} \cdot \textbf{m}^{BFL}\right]
\end{equation}
\begin{equation}
	\varepsilon_{DMI}=D\left( m_z\frac{\partial m_x}{\partial x}-m_x\frac{\partial m_z}{\partial x}+m_z\frac{\partial m_y}{\partial y}-m_y\frac{\partial m_z}{\partial y}\right)
\end{equation}
where, $\sigma=\frac{A_{inter}}{t_{sp2}}$, $A_{inter}$ being inter-layer exchange coupling and $t_{sp2}$ being thickness of spacer-2 region and $D$ is interfacial DMI constant. \\
\indent In the presence of spin current perpendicular to the plane (CPP) of the BFL, the dynamics of skyrmion for layer $l$ of the oscillator is obtained by solving the Landau-Lifshitz-Gilbert-Slonczewski (LLGS) equation described as follows:
\begin{eqnarray}		
	\frac{d\textbf{m}^l}{dt}&=&-|\gamma|\textbf{m}^l\times\textbf{H}^l_{eff}+
	\alpha \left( \textbf{m}^l\times \frac{d\textbf{m}^l}{dt}\right)\nonumber \\
	&&+ |\gamma|\beta_j \epsilon \left( \textbf{m}^l\times \textbf{m}_{RL}\times \textbf{m}^l\right) -|\gamma|\beta_j \epsilon^{\prime}\left(\textbf{m}^l\times\textbf{m}_{RL}\right) 	\nonumber\\		
\end{eqnarray}
where,$\gamma=2.211\times10^5~ \text{m/(A.s)}$ is the gyromagnetic ratio, $\textbf{m}_{RL}$ is the magnetization of RL, $\beta_j=\hbar J/(\mu_0 e t_z M_s)$, $\epsilon=\frac{P\Lambda^2}{\left(\Lambda^2+1\right)+\left(\Lambda^2-1\right)\cdot\left(\mathbf{m}\cdot\mathbf{m}_{RL}\right)}$, and $\epsilon^{\prime}$ is the secondary spin transfer term. Here $e$,$\hbar$, $P$, $t$, and $J$ are electronic charge, reduced Planck's constant, polarization, thickness of the BFL, and current density, respectively. The effective magnetic field $\textbf{H}_{eff}$ is given by:
\begin{equation}\label{Hfield}
	\textbf{H}_{eff}=-\frac{1}{\mu_0 M_s} \frac{\delta E}{\delta \textbf{m}}
\end{equation}
\indent The motion of the center of the skyrmion in the presence of spin current is described by the Thiele equation:
\begin{equation}
	\textbf{G} \times \textbf{v} -\alpha \mu_0 M_s t_z \textbf{v}.\textbf{d}/\gamma + F_{STT} + F = 0
	\label{thiele_eq}
\end{equation}
where the first term represents the Magnus force with the gyro-vector $\textbf{G}$ and velocity of skyrmion $\textbf{v}$. The gyromagnetic coupling vector $\textbf{G}=(4\pi Q\mu_0M_St_z/ \gamma)\hat{z}$ depends on the skyrmion number $Q=-\frac{1}{4\pi} \int dx dy [\textbf{m}\cdot(\partial_x \textbf{m} \times \partial_y \textbf{m})]$, which is equal to $\pm 1$ for an isolated FM skyrmion. The second term represents the dissipative (viscous) force with dissipative tensor $\textbf{d}=\begin{pmatrix}
d & 0\\
0 & d
\end{pmatrix}$, where $d= \frac{1}{4\pi}\int dx dy (\partial_x \textbf{m}\cdot\partial_x \textbf{m})$. The third term represents the force due to spin transfer torque and the last term is the force due to anisotropy gradient (if any) or/and due to boundary-induced force.\\
\indent The driving force due to spin transfer torque and velocity of skyrmion can be decomposed in two parts on the plane of nanodisk $\textbf{F}_{STT}=F_r \hat{r} + F_t \hat{t}$ and $\textbf{v}=v_r \hat{r} + v_t \hat{t}$. The radial force ($F_r$) and tangential force ($F_t$) and are given by $F_{r/t}=-\mu_0 \beta_j M_s t_z \int dx dy[(\textbf{m}\times\textbf{m}_p) \cdot \partial_{r/t}\textbf{m}]$. In the steady state motion of skyrmion on the nano-disk (See Fig.~\ref{oscillator_schematic}(a)) have $v_r=0$ and constant $v_t$ for the given current density. The equation \ref{thiele_eq} can be rewritten along the $\hat{r}$ \& $\hat{t}$ respectively, as:
\begin{eqnarray}\label{thiele_2a}
    v_tG+F_r+F=0 \\ 
    \label{thiele_2b}
    \alpha \mu_0 M_s t_z v_td/\gamma + F_t=0 
\end{eqnarray}
Equation~\ref{thiele_2a} implies that the Magnus force ($v_t$G) is balanced by boundary force or/and anisotropy gradient ($F$). Whereas, the equation \ref{thiele_2b} gives the steady state motion of skyrmion and can be simplified to give $v_t$.  For ease of calculation the $F_t$ during the steady state motion ($v_t=$constant)  can be evaluated when the skyrmion is along the x-axis on nano-disk such that $\textbf{m}_p=-\hat{x}$, $\hat{t}=\hat{-y}$ and $\hat{r}=\hat{x}$.
\begin{equation}
\label{F_t}
F_{t}=-\mu_0 \beta_j M_s t_z \int dx dy[(\textbf{m}\times\textbf{m}_p) \cdot \partial_{y}\textbf{m}] 
%\textbf{m}\times\textbf{m}_p=\hat{y}cos\theta-\hat{z}sin\theta sin\psi
\end{equation}
The profile of skyrmion is given by:
\begin{equation}
\textbf{m}=\hat{x} sin\big(\theta(r')\big) cos \psi + \hat{y} sin\big(\theta(r')\big) sin \psi  + \hat{z} cos\big(\theta(r')\big) 
\end{equation}
where, $\theta(r')$ azimuthal angle is function of $r'$ distance measured from the center of the skyrmion such that $\frac{d\theta}{dr'}\neq0$ only at $r'=R_S$. The $\theta(r')$ changes from $0$ to $\pi$ and $\psi$ changes from $0$ to $2\pi$. The $F_t$ can be simplified the Eq.~\ref{F_t} in the polar coordinate system as:
\begin{equation}
\label{F_t_polar}
F_{t}=-\mu_0 \beta_j M_s t_z \int r' dr'd\psi\Big[sin^2\psi \frac{d\theta}{dr'}+\frac{cos^2\psi  sin2\theta}{2r'}\Big] 
\end{equation}
Assuming $\frac{d\theta}{dr'}\neq0$ only at $r'=R_S$, the Eq.~\ref{F_t_polar} can be further simplified as:
\begin{equation}
\label{F_t_simplifed}
F_{t}\approx-\mu_0 \pi^2 \beta_j M_s t_z R_S
\end{equation}
Using the Eq.~\ref{thiele_2b} \& \ref{F_t_simplifed} the $v_t$ is given by:
	\begin{equation}
		v_t\approx \frac{\gamma \beta_j\pi^2 R_S}{\alpha d}
		\label{v_tan}
	\end{equation}
In the uniform circular motion of skyrmion on the nano-disk the frequency of rotation can be estimated as:
	\begin{equation}
		f^{S}\approx \frac{\gamma \beta_j\pi^2 R_S}{2\pi R_R\alpha d}
		\label{f_T}
	\end{equation}
	where $R_R$ is the radius of rotation of skyrmion in the nano-dot.\\
\indent Thiele equation in the bilayer skyrmion oscillator for an AFC skyrmion pair such that both skyrmions move in unison with the same velocity can be written as:
\\
\begin{eqnarray}		
 \nonumber \textbf{G}_{net} \times \textbf{v} -\alpha^{BFL} \mu_0 M_s t_z \textbf{v}.\textbf{d}^{BFL}/\gamma \\  -\alpha^{TFL} \mu_0 M_s t_z \textbf{v}.\textbf{d}^{TFL}/\gamma 
+ \textbf{F}^{BFL}_{STT} + \textbf{F} &=&M^{SP} a    \label{thielebi_eq}
\end{eqnarray}
where,
$\textbf{G}_{net}=\textbf{G}^{TFL}+\textbf{G}^{BFL}\approx0$ is the vanishing gyro-vector for the AFC skyrmion pair as the total skyrmion number for the top($Q_{TFL}$=1) and bottom free layer($Q_{BFL}$ = -1) is  ($Q_{TFL}$ + $Q_{BFL}$= 0) which means suppressing the SkHE and canceling the Magnus force, $\alpha^{BFL(TFL)}$ and $\textbf{d}^{BFL(TFL)}$ are damping factor and dissipative tensor, respectively of the BFL(TFL). The saturation magnetization and thickness of the BFL and TFL are the same. The vortex spin current is injected only in the BFL which is represented by the $F^{BFL}_{STT}$. $M^{SP}$ is the mass and $a$ is the acceleration of the skyrmion pair.  Assuming that the skyrmion pair is AFC both TFL and BFL skyrmions move together with the same tangential velocity under the influence of  $F^{BFL}_{STT}$, in the steady state motion Eq.~\ref{thielebi_eq} can be decomposed along the $\hat{r}$ \& $\hat{t}$ respectively, as:
\begin{eqnarray}
\label{thielebi_a}
v_tG_{net}+F_r+F=M^{SP}a \\
\label{thielebi_b}
\nonumber \alpha^{BFL} \mu_0 M_s t_z v_td^{BFL}/\gamma \\ + \alpha^{TFL} \mu_0 M_s t_z v_td^{TFL}/\gamma + F_t^{BFL}&=&0 
\end{eqnarray}

STT only acts on the BFL and $F_t^{BFL}$ is given by the Eq.~\ref{F_t_simplifed}. Eq.~\ref{thielebi_b} and can be simplified to give the steady state $v_t$ of the skyrmion pair frequency of oscillation ($f^{SP}$) as:
\begin{eqnarray}
\label{vsp_tan}
v_t\approx \frac{\gamma \beta_j\pi^2 R_S}{\alpha^{BFL}d^{BFL}+\alpha^{TFL} d^{TFL}}\\
		f^{SP}\approx \frac{\gamma \beta_j\pi^2 R_S}{2\pi R_R(\alpha^{BFL} d^{BFL}+\alpha^{TFL} d^{TFL})}
\label{fsp_T}
\end{eqnarray}

We have simulated the proposed device using Object Oriented MicroMagnetic Framework (OOMMF) public code\cite{OOMMF} by incorporating the module for the DMI\cite{DMI_extension}. For the regular single free-layer skyrmion oscillator, the diameter is considered to be 200 nm and the thickness is 0.5 nm. For the bilayer skyrmion oscillator, the diameter for the BFL (TFL) is considered 220 (200) nm and the thickness is 0.5 nm. For both devices, we discretize the free layer into the cell size of 2$\times$2$\times$0.5 $\mathrm{nm^3}$. Material parameters are adopted for the Co/Pt system \cite{Sampaio2013}, which includes exchange stiffness $A=15~ \text{pJ/m}$, Gilbert damping coefficient $\alpha =0.3$,saturation magnetization $M_s=580~\text{kA/m}$, the interfacial DMI constant $D=3~ \mathrm{mJ/m^2}$,polarization $P=0.4$, anisotropy constant $K_u=0.8~\mathrm{MJ/m^3}$, and bilinear surface exchange coefficient $\sigma=-0.25~\mathrm{mJ/m^2}$. For simplicity, we assume $\Lambda$=1, and the secondary spin transfer term $\epsilon^{\prime}$=0.  
\begin{figure}[!b]
		\centering
  \centerline{\includegraphics[width=1\columnwidth]{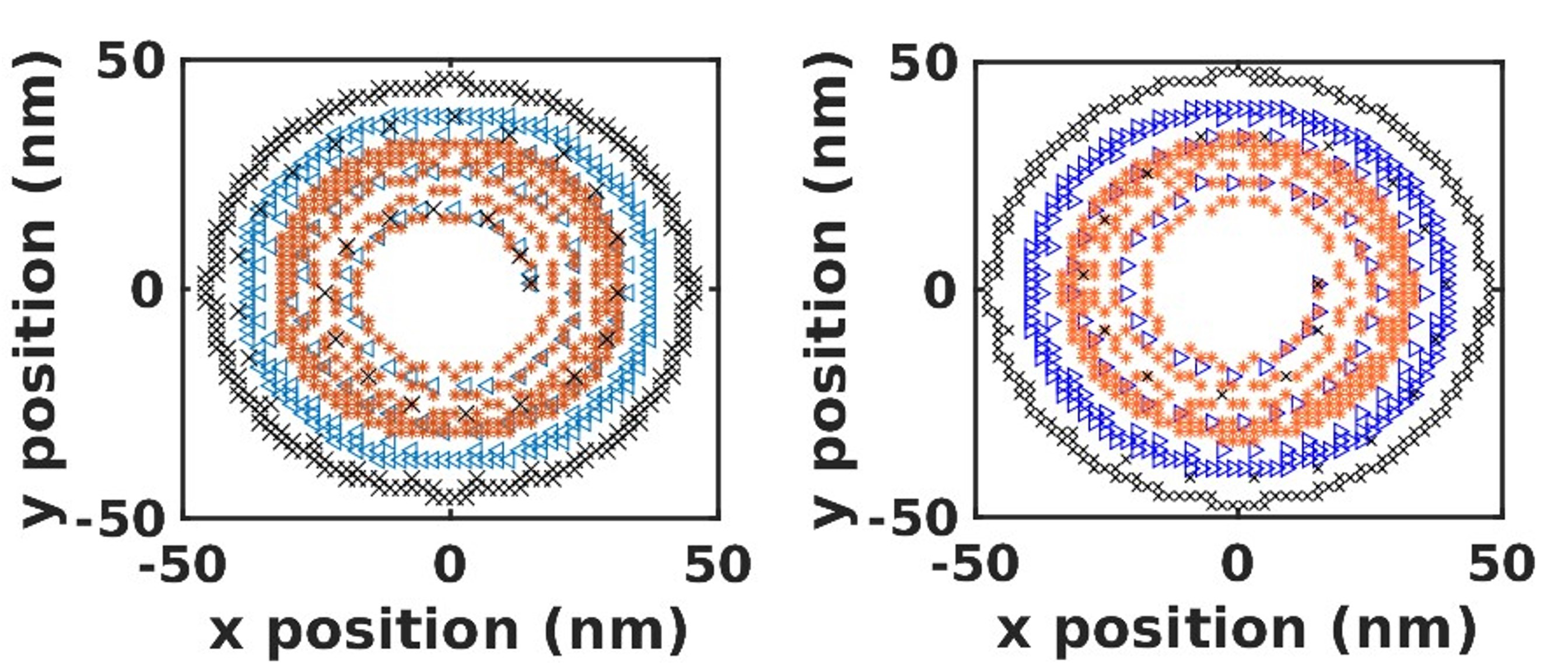}}
 \caption{Single skyrmion(initial:15nm) trajectory in a Bilayer disk with constant Ku profile under the different applied current density of (a)J = -25 $\mathrm{MA/cm^2}$(Orange), -50$\mathrm{MA/cm^2}$(Blue), -100$\mathrm{MA/cm^2}$(Black) for 100ns (b) J = 25 $\mathrm{MA/cm^2}$(Orange), 50 $\mathrm{MA/cm^2}$(Blue),100 $\mathrm{MA/cm^2}$
 (Black) for 100 ns}
\label{bilayer_skyrmion_trajectory}
\end{figure}
\begin{figure}[!t]
		\centering
  \centerline{\includegraphics[width=1\columnwidth]{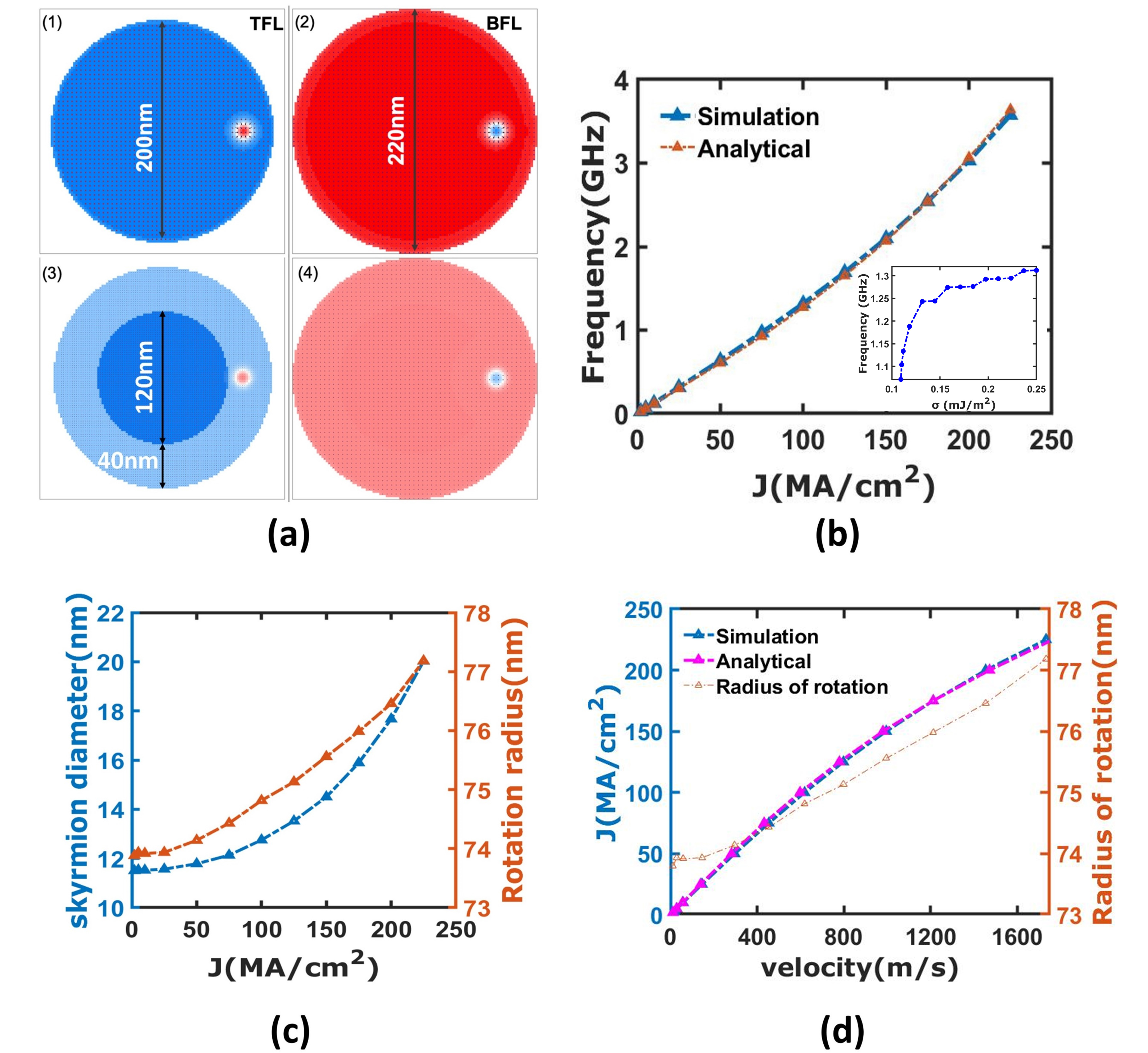}}
		\caption{Bilayer-skyrmion based oscillator. The magnetization of (a.1) the TFL and (a.2) the BFL. The color-coded anisotropy profile of (a.3) the TFL and (a.4) the BFL. The red and blue color represent the magnetization/anisotropy along +$\hat{z}$ and -$\hat{z}$ direction respectively. (b) The frequency of oscillation for both simulation and theoretical obtained from (Eq.~\ref{f_T})along with the bilinear surface exchange coefficient versus frequency in the bilayer skyrmion oscillator at J$= 100~\mathrm{MA/cm^2}$ on the bottom right corner.(c) left y-axis: skyrmion diameter and right y-axis: rotation radius of skyrmion on nano-dot with current density.(d)The Analytical (obtained from (Eq.~\ref{v_tan}) ) and simulated velocity of skyrmion at different current densities and their corresponding radius of rotation for the bilayer skyrmion nano-oscillator}
		\label{bilayer_skyrmion_oscillator_chracterstics}
	\end{figure}
\section{Results and discussion}
\label{results}
We reported simulations on both monolayer and bilayer oscillators to investigate the impact of Magnus force on skyrmions. The device assumes that the initial magnetization of the FL without the skyrmion is oriented towards $+z$-direction. Previously reported result\cite{garcia2016skyrmion} shows that a skyrmion at the center of the nanodisk remains immobile even after injecting the spin current. Thus for the monolayer oscillator (Fig. \ref{oscillator_schematic}), we first nucleate a skyrmion on the FL at a distance of 25 nm from the center of the nanodisk as shown in Fig. \ref{skyrmion_oscillator_chracterstics}(a). Then, we apply the current density (J) with the spin polarization according to $\textbf{m}_{\text{RL}}$ into the FL. We start with a value of J=2 $\mathrm{MA/cm^2}$, which initially generates a force on the skyrmion along radial ($F_r$) as well as tangential ($F_t$) direction, and eventually the skyrmion moves towards the edge of the nanodisk. Due to the tilted magnetization at the edge of the nanodisk, it exerts a repulsive force along the radially inward direction. When the skyrmion reaches near the edge, this repulsive force ($F_b$) counterbalances $F_r$, resulting in a stable circular motion of the skyrmion on the nanodisk as shown in Fig. \ref{skyrmion_oscillator_chracterstics}(a)\cite{Videolink} with a frequency of $\sim$45 MHz (Fig. \ref{skyrmion_oscillator_chracterstics}(b)). With the increase of $J$, the frequency of this oscillation also increases as shown in Fig. \ref{skyrmion_oscillator_chracterstics}(b). This is justified, as the velocity of the skyrmion increases with the injected current density\cite{Sampaio2013,Das2019}, which eventually increases the frequency, and Eq. \ref{f_T} also supports this claim.
With the increase of $J$, the tangential velocity also increases (Eq.~\ref{v_tan}), which further increases the Magnus force acting on the skyrmion. This pushes the skyrmion toward the nanodisk edge. This is evident from Fig. \ref{skyrmion_oscillator_chracterstics}(c), which shows the radius of the circular path ($\mathrm{R_{rot}}$) traversed by the skyrmion increases with $J$. As the skyrmion moves toward the edge of the nanodisk, the edge repulsion force becomes more dominant, which acts against the radial force felt by the skyrmion due to the injected current. This shrinks the size of the skyrmion as we can see from Fig. \ref{skyrmion_oscillator_chracterstics}(d) that shows the decrease of the skyrmion diameter with an increasing $J$. We have obtained the highest frequency of 120 MHz at $J$=7 $\mathrm{MA/cm^2}$. Further increment of $J$ increases the velocity, which generates a large Magnus force that pushes the skyrmion to the edge of the nanodisk, and eventually annihilates the skyrmion.  \\
\indent Similarly, for the case of a bilayer skyrmion oscillator having a constant Ku profile in both layers the skyrmion moves radially outward in the anti-clockwise direction (Fig.\ref{bilayer_skyrmion_trajectory}a)(clockwise direction(Fig.\ref{bilayer_skyrmion_trajectory}b)) for the different applied current densities in 100ns simulation time depicting skyrmion motion to be unaffected by the current magnitude. Now, we design the bilayer skyrmion oscillator with different uniaxial anisotropy profiles for enhanced frequency operations.

\begin{figure}[!b]
		\centering
          \centerline{\includegraphics[width=1\columnwidth]{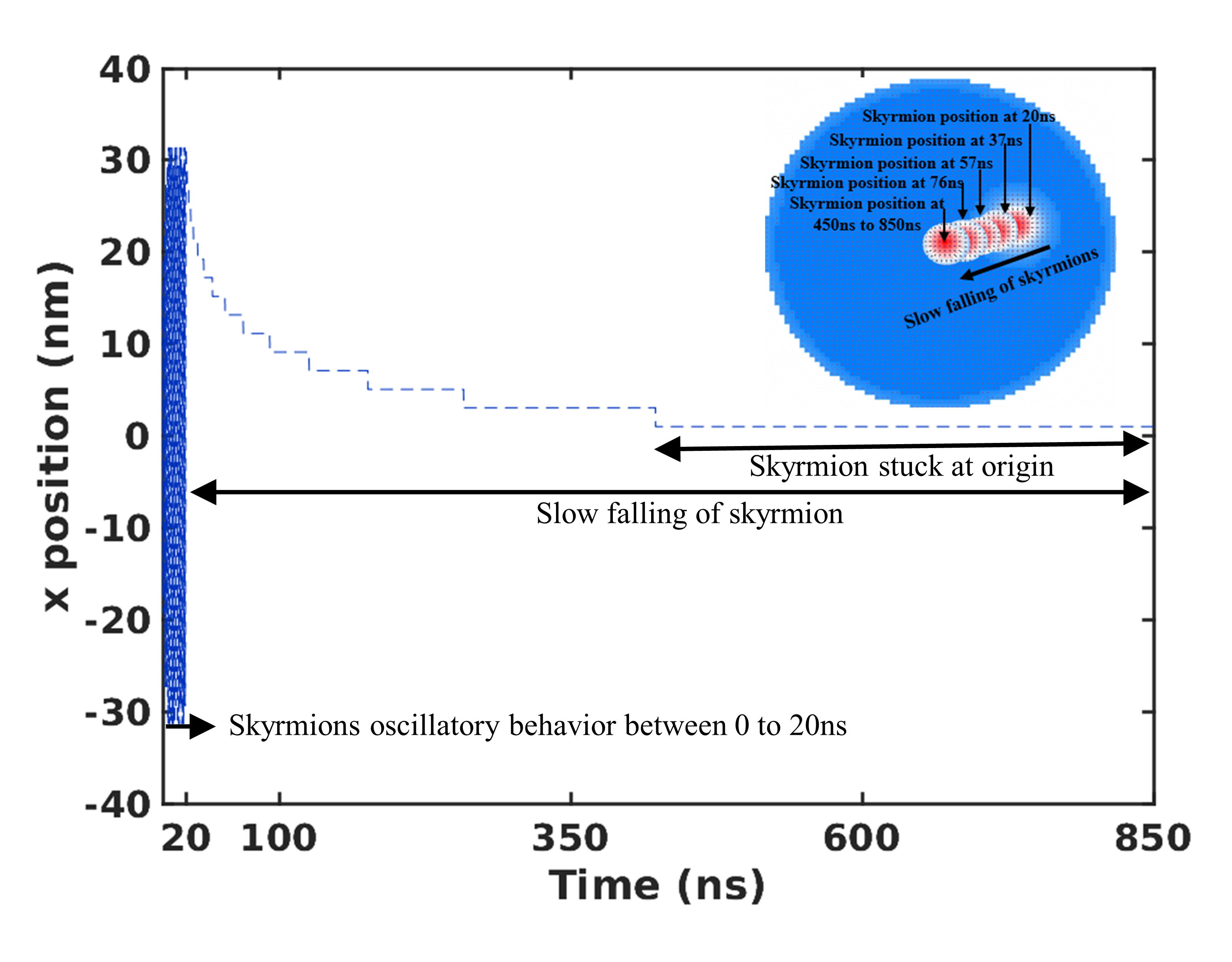}}
	\caption{The plot shows the x-position of the skyrmion's center over time, illustrating its oscillatory motion before 20 ns with current and its slow falling towards the nano-disk center from 20 ns to 850 ns without current. Snapshots of the skyrmion's position at different time instants are included. The simulation uses a current density of J=25 ~$\mathrm{MA/cm^2}$ and runs for 850ns.}
	\label{slowfalling_combined}
     \end{figure}

\indent We show in Fig.~\ref{bilayer_skyrmion_oscillator_chracterstics}(a.1 \& 2)\cite{Videolink} magnetization of the TFL and BFL of the bilayer skyrmion oscillator. 
We show in fig.~\ref{bilayer_skyrmion_oscillator_chracterstics}(a.3 \& 4) the anisotropy profile of the TFL and BFL. We have used a high anisotropy disc in the TFL of diameter 120 nm (see Fig.4 a(3)) to avoid the slow falling of the skyrmion pair.Slow falling refers to the gradual movement of the skyrmion towards the center of the nano-disk until equilibrium in the absence of current. This behavior is predominantly due to the edge repulsion forces, with no driving force from the applied current. The skyrmion gets trapped at the center\cite{garcia2016skyrmion}, making it difficult to restart oscillation even after injecting the spin current. \\To prevent this, a high anisotropy disc is added to the TFL. Fig.\ref{slowfalling_combined} illustrates the skyrmion's x-position and snapshots at various time instants, showing its slow fall between 20 ns and 850 ns without applied current density.
The vortex-like spin-polarized current is injected in the BFL, which moves the AFC skyrmion pair.
We show in Fig.~\ref{bilayer_skyrmion_oscillator_chracterstics}(b) the  theoretical(Analytical) curves of the frequency of oscillation for the different current densities in the bilayer skyrmion oscillator derived from (Eq.~\ref{f_T}) with a solid orange curve is in very good agreement with the simulation curve shown with a solid blue curve.The absence of Magnus force ($G=0$) in the AFC skyrmion pair allows the large injection of current density that makes the bilayer skyrmion oscillator a suitable candidate for generating high-frequency oscillations. We also observe the AFC skyrmion-pair diameter also grows at large vortex spin current densities as shown in Fig.~\ref{bilayer_skyrmion_oscillator_chracterstics}(c). The skyrmion pair in the bilayer skyrmion oscillator can not enter in the high anisotropy region ring, which results in the expansion of the skyrmion center towards the edge of the nano-disk. This increases the skyrmion rotation radius as shown in Fig.~\ref{bilayer_skyrmion_oscillator_chracterstics}(c). This design works for a bilinear surface exchange coefficient values of above -0.1090 mJ$/m^2$ (Fig.~\ref{bilayer_skyrmion_oscillator_chracterstics}(d))  \\

\begin{figure}[t!]
		\centering
  \centerline{\includegraphics[width=1\columnwidth]{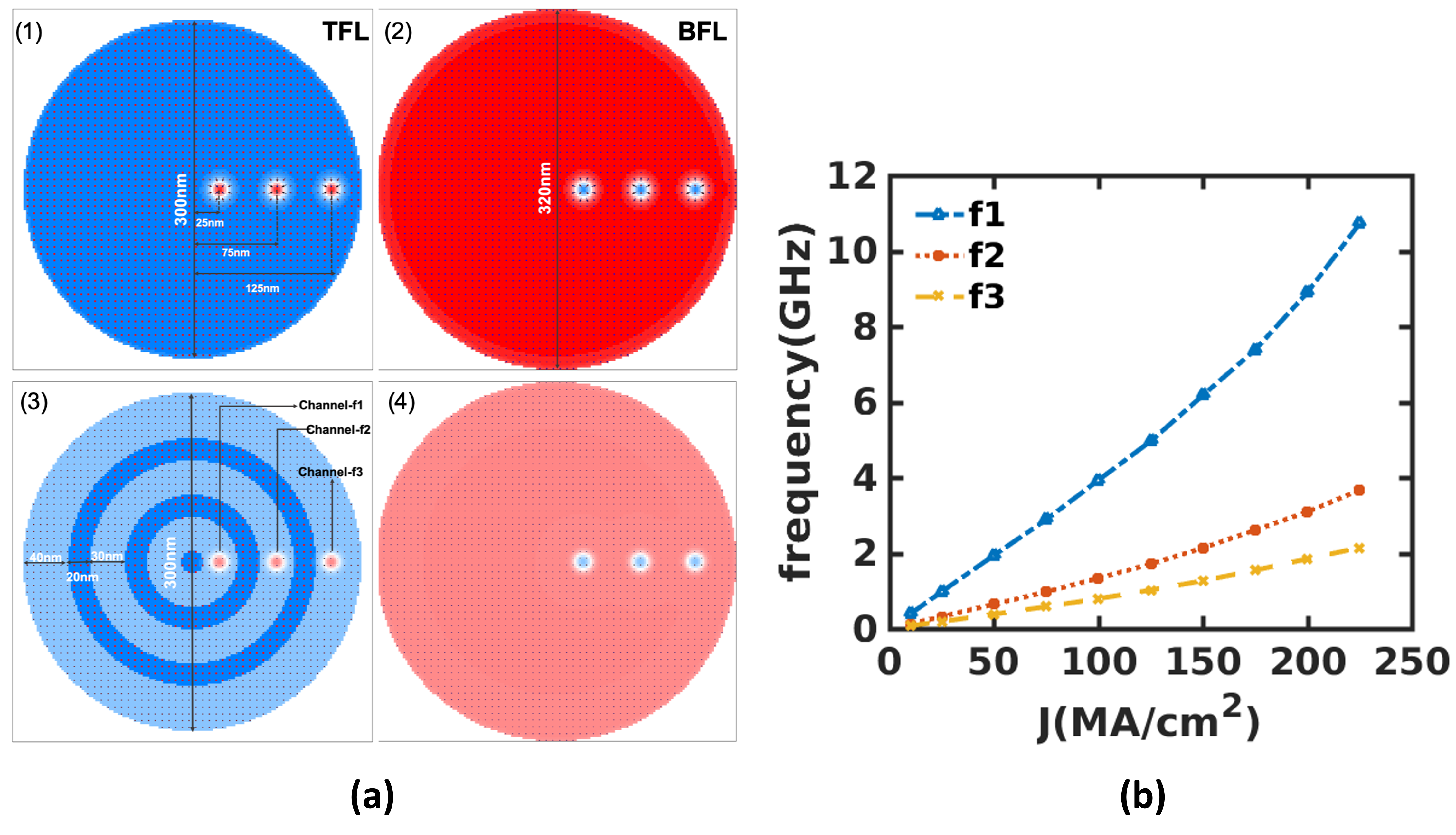}}		
		\caption{Multi-channel bilayer-skyrmion based oscillator. The magnetization of (a.1) the TFL and (a.2) the BFL. The color-coded anisotropy profile of (a.3) the TFL and (a.4) the BFL. (b) The frequency of oscillation.}
		\label{bilayer_skyrmion_oscillator_multi_channel}
	\end{figure}

\indent The absence of the Magnus force in the bilayer skyrmion oscillator paves a pathway for a multi-channel frequency oscillator excited by the same current density. The tangential velocity (Eq.~\ref{vsp_tan}) of the skyrmion-pair in the bilayer skyrmion oscillator depends on the current density.
As shown in Fig.\ref{bilayer_skyrmion_oscillator_chracterstics}(d) both the analytical and simulation curves are in accord. Also, we observe a very minute change of 4-5nm in the radius of rotation of the skyrmion for the different current densities. thus, velocity is weakly dependent on the radius of rotation.The skyrmion pairs at different radial distances from the center of the nano-disk under the same current density have the same tangential velocities and different angular velocities. This can be used to make a multi-channel oscillator driven by the same current source. We present the design of a 3-channel oscillator shown in Fig.~\ref{bilayer_skyrmion_oscillator_multi_channel}\cite{Videolink} out of a large landscape of designs for such multi-channel oscillators. The physical dimensions of the bilayer nano-disk for the multi-channel oscillator are shown in Fig.~\ref{bilayer_skyrmion_oscillator_multi_channel}(a) such that the skyrmion-pairs in two different channels do not reflect each other. The centripetal force by the effective mass of the AFM skyrmion in the bilayer device with channel-f3 gyrates the skyrmion in the absence of Magnus force. In channels f1 and f2, skyrmions are bound by the repulsive forces of the high Ku nanorings, ensuring balance and stability in the formed spaces by PMA patterning\cite{juge2021helium} between the high Ku
nanorings.The high anisotropy energy barrier region in the TFL separates the skyrmion pairs of different channels. It also provides robustness to the motion of the skyrmion pairs against any perturbation in their respective channels. We show in Fig.~\ref{bilayer_skyrmion_oscillator_multi_channel}(b) the frequency of oscillation for the channel-1, channel-2, and channel-3 of the oscillator. The Channel-1 has the highest frequency ($\approx$11GHz) due to a smaller radius of rotation(Eq.~\ref{fsp_T}).\\
 
 \begin{figure}[h!]
		\centering
  \begin{center}
  \centerline{\includegraphics[width=1\columnwidth]{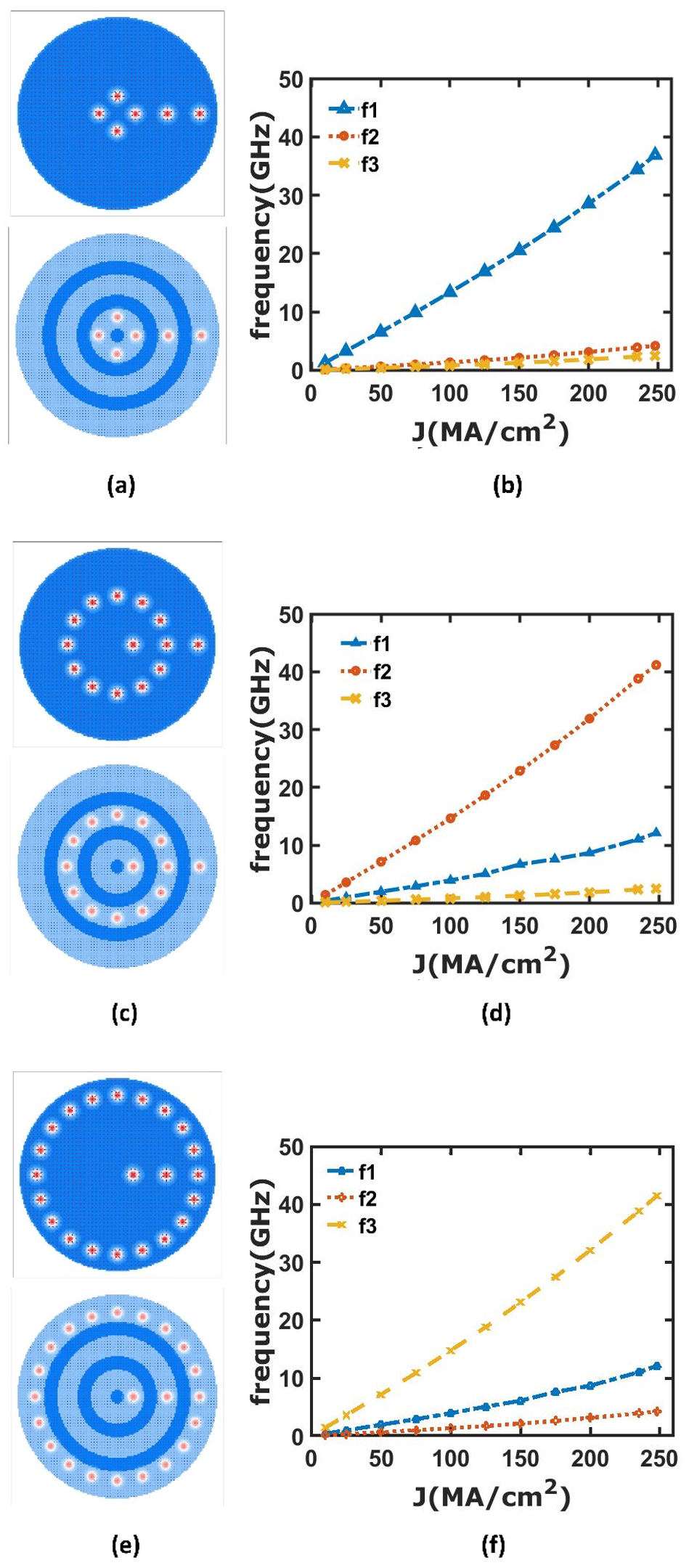}} 
          \end{center}
		\caption{Multi-channel bilayer-skyrmion based oscillator with multi-skyrmion (a) The magnetization and anisotropy profile of the TFL in channel-1.(b)The frequency of oscillation of the different channels in channel-1.(c)The magnetization and anisotropy profile of the TFL in channel 2.(d) The frequency of oscillation of the different channels in channel 2. (e)The magnetization and anisotropy profile of the TFL in channel-3 (f) The frequency of oscillation of the different channels in channel-3.}
		\label{bilayer_multi_skyrmion_oscillator_multi_channel_1}
	\end{figure}
The frequency of the multi-channel bilayer skyrmion oscillator can be increased further by nucleating multiple skyrmions in the channel as now the presence of the next skyrmion in the skyrmion lattice shadows the current one, resulting in a shorter time period and higher frequency generation compared to using a single skyrmion in the modified design. 
\indent The frequency of the multi-channel bilayer skyrmion oscillator can be increased further by nucleating multiple skyrmion in the channel. We show in   Fig.~\ref{bilayer_multi_skyrmion_oscillator_multi_channel_1}(a)\cite{Videolink} the magnetization profile of the TFL of the multi-channel skyrmion oscillator with 4-skyrmion pairs in channel-f1. The maximum number of skyrmion pairs that can be generated in channel-1 is given by $N_s\approx\frac{2\pi R_R }{4R_S}$. We have 4-skyrmion pairs in the channel-1 (Fig.~\ref{bilayer_multi_skyrmion_oscillator_multi_channel_1}(a)) which increases the frequency of oscillation of the channel-1 by a factor of 3.4 as shown in the Fig.~\ref{bilayer_multi_skyrmion_oscillator_multi_channel_1}(b). The number of skyrmion pairs in a particular channel should be maximum such that different skyrmion pairs are interlocked due to skyrmion-skyrmion repulsion. 
The constricted high Ku rings that form the nano-tracks motivate the skyrmion interlocking secured by skyrmion-skyrmion repulsion. This leads to single-frequency output and robustness against deflection of skyrmion motion due to scattering. \\
\indent The frequency of channel-2 can also be increased by using multi-skyrmions in channel-2. We show in Fig.~\ref{bilayer_multi_skyrmion_oscillator_multi_channel_1}(c)\cite{Videolink} multi-skyrmions in channel-2 with a maximum of 12 skyrmion pairs. In our design of the multi-channel skyrmion oscillator, the radius of rotation of channel-2 skyrmion-pair is three times that of channel-1 (see Fig.~\ref{bilayer_multi_skyrmion_oscillator_multi_channel_1}(b)). It allows the maximum generation of 12 skyrmion pairs in channel 2. The frequency of oscillation of channel 2 is shown in Fig.~\ref{bilayer_multi_skyrmion_oscillator_multi_channel_1}(d). It can be noted that the maximum frequency ($\approx$41 GHz) of channel-2 with multi-skyrmion pairs is increased to that of channel-1($\approx$37 GHz) with multi-skyrmion pairs (Fig.~\ref{bilayer_multi_skyrmion_oscillator_multi_channel_1}(d) \& Fig.~\ref{bilayer_multi_skyrmion_oscillator_multi_channel_1}(b)).\\
\indent  We show in  Fig.~\ref{bilayer_multi_skyrmion_oscillator_multi_channel_1}(e)\cite{Videolink} multi-skyrmions in channel-3. The radius of rotation of channel-3 skyrmion-pair is five times that of channel-1 (see Fig.~\ref{bilayer_multi_skyrmion_oscillator_multi_channel_1}(b)). In channel 3, we can generate a maximum of 20 skyrmion pairs. The multi-skyrmion increases the frequency of channel-3 as shown in Fig.~\ref{bilayer_multi_skyrmion_oscillator_multi_channel_1}(f). It has to be noted that the maximum frequency ($\approx$41 GHz) is the same as that of channel 2. In all three profiles with multi-skyrmion, the spacing between the neighboring skyrmions remains nearly constant, with only a small difference of approximately 2.2 nm. This small variation contributes to the nearly identical high frequencies observed in these profiles.

\section{Conclusion}
\label{conclusion}
\indent The paper presents a multichannel SAF skyrmion-based STT nano-oscillator design capable of generating ultrahigh microwave frequencies in the broad range of 0-41GHz. Such enhanced frequency range is achieved by eliminating the Magnus force using a bilayer device as well as design modification considering the Ru spin-flip length by incorporating a reduced spacer region that minimizes the spin current flow in the TFL. We further demonstrate the tunability of the frequencies based on the number of skyrmions interlocked in the multichannel high anisotropy rings for the same current density variation. We conclude that the multi-channel bilayer oscillator with 1,12,1 skyrmions in their respective channels has an enhanced frequency range(0-41GHz) as compared to various other designs such as the monolayer skyrmion oscillator(max$\sim$120MHz), A bilayer skyrmion oscillator with a high Ku disc(max$\sim$3.8GHz) and a multi-channel bilayer oscillator with one skyrmion per channel (max$\sim$10GHz) are demonstrated in the paper. Finally, we conclude that this work provides a generic method to generate multi-channel frequency output along with ultra-high frequency in a controllable manner which may open new pathways to utilize skyrmion-based nano-oscillators in various applications.
\printbibliography
\nocite{*}
\end{document}